# Self-Consistent Disk-Reflection Analysis of the Black-Hole Candidate X-ray Binary MAXI J1813-095 with *NICER*, *Swift*, *Chandra*, and *NuSTAR*

Santiago Ubach,[1,2] James F. Steiner,[1] Jiachen Jiang,[3] Javier García,[4,5] Riley M. T. Connors,[6] Guglielmo Mastroserio,[7] Ye Feng,[8] and John A. Tomsick[9]

[1] *Harvard-Smithsonian Center for Astrophysics, 60 Garden Street, Cambridge, MA 02138, USA*
[2] *Departament de Física & CERES-IEEC, Universitat Autònoma de Barcelona, Bellaterra, Spain*
[3] *Department of Physics, University of Warwick, Gibbet Hill Road, Coventry CV4 7AL, UK*
[4] *X-ray Astrophysics Laboratory, NASA Goddard Space Flight Center, Greenbelt, MD 20771, USA*
[5] *Cahill Center for Astronomy and Astrophysics, California Institute of Technology, Pasadena, CA 91125, USA*
[6] *Department of Physics, Villanova University, 800 E Lancaster Ave, Villanova, PA 19085, USA*
[7] *Dipartimento di Fisica, Università degli Studi di Milano, Via Celoria, 16, Milano, 20133, Italy*
[8] *Key Laboratory for Computational Astrophysics, National Astronomical Observatories, Chinese Academy of Sciences, Datun Road A20, Beijing 100012*
[9] *Space Sciences Laboratory, 7 Gauss Way, University of California, Berkeley, CA, 94720-7450, USA*

## ABSTRACT

We present our analysis of MAXI J1813-095 during its hard state "stalled" outburst in 2018. This self-consistent analysis has been carried out using *NICER*, *Swift*, *Chandra*, and *NuSTAR* throughout seven observations of MAXI J1813-095. We find a relativistic iron line at ∼6.5 keV from the inner region of the accretion disk. Our results are consistent with a slightly truncated disk or non-truncated disk for an inner radius of ∼$2R_\mathrm{g}$ and minimum spin of >0.7 with a best value of ∼ 0.9, assuming $R_\mathrm{in}$ reaches the innermost stable circular orbit at $L_\mathrm{x} \sim 1\%\, L_\mathrm{Edd}$. We analyzed MAXI J1813-095 over its outburst employing a spectral model which self-consistently couples the seed disk photons to the Comptonization and reflection components, also inclusive of reflection Comptonization. The unique aspect of this work is a reflection fraction of order unity, which is significantly higher than previous studies of this source, and is a consequence of applying the self-consistent disk-Comptonization-reflection spectral model. Other key parameters such as inclination and inner radius are found to be consistent with other works.

## 1. INTRODUCTION

The most extreme stellar compact objects, black holes (BHs) and neutron stars (NSs), are often revealed when they reside in X-ray binary (XRBs) systems. Black hole X-ray Binaries (BHXRBs) are bright and variable X-ray emitters found in the Milky Way and nearby galaxies (e.g.Hertz & Grindlay 1983; Jordán et al. 2004; Kalemci et al. 2022). The X-rays emitted by these compact objects are produced by a disk of accreting gas supplied by the outer layers of its companion star, either from Roche-lobe overflow or through stellar wind capture. Depending on the companion mass the system is classified as either a low-mass or high-mass X-ray binary (e.g. Liu et al. 2007, 2006).

Low-mass BHXRBs commonly undergo transient outbursts beginning with a sudden and rapid increase of the disk accretion rate and corresponding brightening of X-ray luminosity. During outbursts, two broad, characteristic spectral states ("soft" and "hard") are distinguished (see Remillard & McClintock 2006 & Belloni 2010). In the soft state, the X-ray emission is dominated by thermal emission from the accretion disk and in hard states, the dominant emission is non-thermal which arises from a hot corona enshrouding the disk. The hard state is also associated with the presence of a steady jet (e.g., Cui et al. 1998; Rao et al. 2000; Fender 2001; Yu & Yan 2009; Shidatsu et al. 2011; Reis et al. 2013). It is common to separate transitional states as "intermediate", sometimes further distinguishing as "hard-" or "soft-intermediate" (e.g. Homan & Belloni 2005 & Belloni 2010).

These transients evolve in their spectral and timing properties during outbursts. Major outbursts exhibit state transitions, but some outbursts, which often have both lower X-ray fluence and lower peak luminosity, do not (Alabarta et al. 2021).

Such cases are commonly termed as "failed", "stalled" or in the convention of Debnath et al. (2017), "Type 2" outbursts. Whereas in major outbursts (Debnath et al. 2017 "Type 1"), BHXRBs exhibit the following canonical sequence of spectral states: hard, hard-intermediate, soft-intermediate, soft, followed by a lower-luminosity transition through the intermediate and then hard state, completing the full outburst cycle. When plotted on a hardness-intensity diagram (HID), this traces out a "q"-shaped pattern (e.g. Fender et al. 2004; Homan et al. 2005; Dunn et al. 2010).

The 'no-hair theorem' states that an astrophysical BH is completely describable by just two quantities: mass and spin (for a review, see Chruściel et al. 2012; Bambi 2018; Reynolds 2021). Measurements of BH spins are important for the role spin may play in powering jets or in governing accretion disk dynamics (e.g., McClintock et al. 2014). There are two primary spectroscopic techniques to measure BHXRB spins, the X-ray continuum fitting method (Zhang et al. 1997) and the relativistic reflection method (Fabian et al. 1989). Both rely on the relationship between the BH spin and the inner-disk radius, which is linked to the innermost stable circular orbit (ISCO). For example, quasi-periodic oscillations (QPOs) (e.g., Stella & Vietri 1998; Stella & Vietri 1999; Čadež et al. 2008; Kostić, U. et al. 2009; Bakala et al. 2014; Karssen et al. 2017; Germanà 2017), gravitational waves (GWs) (e.g.,Ajith et al. 2011; Santamaría et al. 2010; Vitale et al. 2017), and X-ray polarimetry are capable of offering spin constraints (e.g., Connors et al. 1980; Dovčiak et al. 2008; Li et al. 2009). X-ray polarimetry can also constrain the inclination and coronal geometry (e.g., Krawczynski et al. 2022).

An active area of research is whether the disk extends to the ISCO throughout its outburst or whether it is truncated at $R_{in} > R_{ISCO}$ in certain conditions (e.g., García et al. 2015; Kostić, U. et al. 2009). In particular, there is a debate as to whether the inner accretion disk of a BH X-ray binary is truncated in the hard states e.g., as suggested in Esin et al. (1997).

The BH candidate MAXI J1813-095 (hereafter MAXI J1813) was discovered on 2018 February 19 (Kawase et al. 2018) with *MAXI*/GSC. Follow up observations with the *Swift*/XRT localized the source to RA = $18^h13^m34.0^s$; DEC = -09°31′59.0″ (Kennea et al. 2018). GROND followup identified an optical counterpart (Rau 2018), and ATCA observations revealed a compact jet and classified the source as a likely radio-quiet BHXRB (Russell et al. 2018). From multi-wavelength observations, Armas Padilla et al. (2019) suggested that the companion star could be a G5V star with a distance of > 3 kpc. While the nature of the compact object in MAXI J1813 has not yet been dynamically confirmed, its characteristics are consistent with a BH primary and it is not known to have exhibited pulsations or thermonuclear X-ray bursts.

In this work, we study the geometry of the inner accretion flow of MAXI J1813 in its hard state via a self-consistent consideration of coronal Comptonization in reflection spectroscopy. As we demonstrate, its X-ray emission is consistent with being dominated by non-thermal coronal power-law emission and associated reflection features, as is typical of a BH transient in the hard state (e.g. Fuerst et al. 2018; Armas Padilla et al. 2019).

Figure 1 shows the 2018 MAXI lightcurve in X-rays of MAXI J1813 of the full outburst which lasted ~90 days. The stalled outburst as seen in 2018 for MAXI J1813, quickly reached the peak of X-ray flux within the first observations of its outburst. The observations with *NICER*, *Swift*, *NuSTAR* and *Chandra* are shown as solid vertical lines, corresponding to each observation. In the cases where the observations didn't overlap, multiple lines are drawn in Figure 1.

In this study, we analyze seven observations of MAXI J1813 with four different instruments. Five observations use *NICER*, all seven use *Swift*, three observations use *Chandra* and three observations use *NuSTAR* (see Table 1). Hereafter, these will be referred to as Obs 1–7. We focus on a self-consistent reflection spectroscopic analysis of MAXI J1813's hard state in which we examine the system properties and consider the question of disk truncation.

X-ray reflection in XRBs involves the reprocessing of high-energy X-rays produced in the corona at the accretion disk surface, leading to the characteristic signatures (e.g., Fe-K$\alpha$ fluorescence) which can be observationally constrained. Jiang et al. (2022) studied MAXI J1813 using the last 3 observations in our analysis including data from *Swift* and *NuSTAR*. They fit the data using a variety of relativistic disk reflection and the Comptonization models. The three reflection models included an earlier version of "*relxillCp*" which keeps the electron temperature fixed at 300 keV. The other two reflection models used in their analysis are "*relxilllpcp*" (Dauser et al. 2016) and "*reflionx*" (Ross & Fabian 2005). They analyzed each of the three observations with all three models, using *nthcomp* for the Comptonization of the disk (Życki et al. 1999) and *tbnew* to account for Galactic absorption (Wilms et al. 2000). A continuum-only joint analysis to assess the prominence of reflection features, showed a clear Fe-K emission line at ~6.5 keV and a Compton hump at ~20 keV with a stable $\Gamma$ ~1.6-1.7. From the reflection analysis, The iron abundances were near solar, and the ionization was high (~3). The best joint fit was obtained using "*relxilllpcp*". From this, the reflection fraction was found to be low, ~0.18; notably this value was significantly lower, ~ 0.08 in the fit with *relxillCp*, owing to those models respective geometric assumptions regarding the corona.

Jana et al. (2021) studied MAXI J1813 using *NICER*, *NuSTAR* and *Swift* data but using less data than we consider in our analysis. Jana et al. (2021) performed a timing analysis with the *NICER* data finding no quasi-periodic oscillations (QPOs). Their spectral analysis was performed using a disk blackbody component and a power-law model to extract the thermal and non-thermal fluxes, and to find the photon index and inner disk temperature. They included Galactic absorption (*TBabs* (Wilms et al. 2000)) and a gaussian to fit the Fe-K, with a full model *TBabs(diskbb + power-law + gaussian)*. Another model they explored was *reflect* model in order to determine the reflection fraction, photon index and inclination. Lastly, they used a physical two-component advective flow (TCAF) model to estimate the mass of the black hole and a *laor* component to model the Fe-K$\alpha$ reflection line. The estimated black hole mass was found to be $\sim 7.4 M_\odot$.

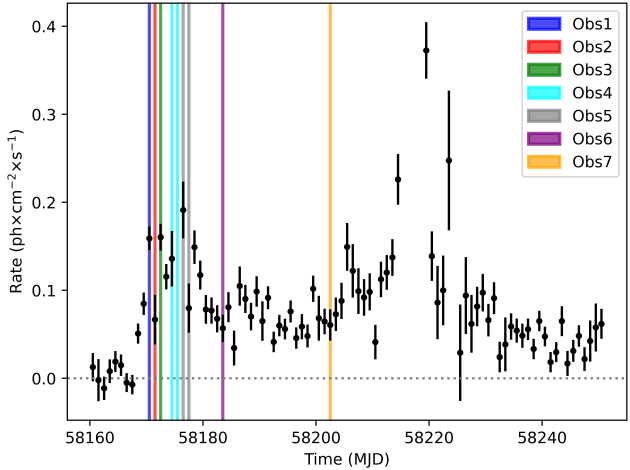

**Figure 1.** The lightcurve of MAXI J1813 form the MAXI telescope during the stalled outburst from 2 to 20 keV. MAXI J1813 reaches its peak of luminosity quickly. The vertical solid lines, correspond to each observation (see Table 1). Observations 4 and 5 were not taken strictly simultaneously by each observatory, but were within 1 day. We therefore mark them as two separate lines in each case.

In Section 2, we describe the observations and data reductions. In Section 3, we present the data and our analysis utilizing a self-consistent spectral reflection model. In Section 4 we discuss our results across the different observations. In Section 5, we present our conclusions.

## 2. DATA SELECTION AND REDUCTION

We used data from four different instruments. We aggregate observations within ~24 hours of one another among these instruments to maximize coverage of the source. We took data from 21-02-2018 to 25-03-2018 (see Table 1) where *Swift* is present in all of the observations overlapping with *NICER* in Obs 1–5 and with *Chandra* and *NuSTAR* in Obs 5–7. *NICER* and *Swift* energy ranges were chosen in such a way as to ensure that the data was uniformly above the background. Although we have aggregated the observations, in some cases there may be more than one distinct good-time-intervals (GTIs) for a given instrument's observation. In such instances, those GTIs are treated as distinct spectra.

### 2.1. *NICER*

*NICER*, the Neutron star Interior Composition Explorer is a space-based X-ray instrument focused on soft X-ray timing/spectroscopy (Gendreau et al. 2012) with an energy range of 0.2 - 12 keV in which we fitted over 0.7-10 keV.

*NICER* observations were first screened using default settings in NICERDAS v8, i.e., requiring distance from the Earth limb > 15°, distance from the *bright* Earth limb > 30°. Data from the South-Atlantic Anomaly are excluded, as well as any data within 60° of the Sun unless the "sunshine" flag is equal to 0. The individual focal-plane modules (FPMs) were screened to remove any outliers on the basis of undershoot, overshoot, or X-ray event rates with a $\sim 10\sigma$ threshold of variance with respect to the detector ensemble for rejection. Detectors 14, 34, and 54 were excluded from analysis owing to potential calibration issues with these detectors. The resulting data products were screened for edge-clipping particle events using the standard "trumpet filter" on the "PI ratio" (i.e., the ratio of slow-chain to fast-chain event energies), and the 3C50 background model was employed for spectral analysis (Remillard et al. 2022). Response products have been generated using calibration release xti20210707[1].

### 2.2. *Swift*

The *Swift* observatory is a multi wavelength facility with the capacity to observe from ultraviolet/optical wavelengths (UVOT) to X-rays (XRT) and $\gamma$-rays (BAT). For our analysis we only used the XRT data (Burrows et al. 2005) in the range of 0.7-10 keV.

*Swift* observed MAXI J1813 at three epochs simultaneously with *NuSTAR* and *Chandra*, and five epochs with *NICER* over a total of seven observations. In total, *Swift* observed MAXI J1813 twelve times between 2018 February 20 and 2018 March 25, of which we only used those coinciding with observations including other instruments. *Swift*/XRT observations of MAXI J1813 were carried out in windowed-timing (WT) mode. We extracted cleaned event files with the FTOOLS task xrtpipeline version 0.13.7 and calibration (CALDB) file version 20210915. We chose a box region of radius 95"×90" for source and 180"×90" for the background. Light curves, source and background spectra were extracted by using XSELECT version 2.5a.

### 2.3. *NuSTAR*

*NuSTAR* (Harrison et al. 2013) is the first hard X-ray focusing observatory launched by NASA. It consists of two identical focusing modules: FPMA and FPMB. We reduced the *NuSTAR* data using the *NuSTAR* Data Analysis Software (NuSTARDAS) package and calibration data of v20210824. The spectra of MAXI J1813 were extracted for both FPMA and FPMB detectors from a 100" radius circle centered on the source, while the background spectra were extracted from a source-free circle of 200" on the same detector chip. Data are analyzed from 3-75 keV for each FPM.

---

[1] We have confirmed that differences are negligible within statistical uncertainty using the most recent calibration release (20221001).

**Table 1.** Log of *NICER* (Ni), *Swift* (S), *Chandra* (Ch), and *NuSTAR* (Nu) observations of MAXI J1813-095.

| Observations | Data Sets | Date (mm-dd-yyyy) | Obs. ID | Exposure times (ks) |
|---|---|---|---|---|
| Obs 1 | Ni1+S1 | 02-21-2018 | 1200090101 & 00010563002 | 0.6 & 1 |
| Obs 2 | Ni2+S2 | 02-22-2018 | 1200090102 & 00811167000 | 0.4 & 0.25 |
| Obs 3 | Ni3+S3 | 02-23-2018 | 1200090103 & 00010563003 | 2.4 & 0.9 |
| Obs 4 | Ni4+S4 | 02-25/26-2018 | 1200090104 & 00010563004 | 1.1 & 0.4 |
| Obs 5 | Ni5+S5+Ch1+Nu1 | 02-27/28-2018 | 1200090105 & 00088654001 & 20264 & 80402303002 | 1.3 & 1.9 & 20 & 23.2 |
| Obs 6 | S6+Ch2+Nu2 | 03-06-2018 | 00088654002 & 20265 & 80402303004 | 1.8 & 19.9 & 20.5 |
| Obs 7 | S7+Ch3+Nu3 | 03-25-2018 | 00088654003 & 20266 & 80402303006 | 1.9 & 18.5 & 20.4 |

### 2.4. *Chandra*

*Chandra* (Weisskopf et al. 2000) is the premier soft X-ray imaging facility, and has been operating for 25 years. We make use of *Chandra*'s Advanced CCD Imaging Spectrometer (ACIS) instrument. Owing to a contamination layer on the filter, ACIS observations have reduced sensitivity below $\sim 1-2$ keV.

*Chandra* observed MAXI J1813 during three epochs coinciding with *Swift* and *NuSTAR* (Obs 5–7), with *NICER* data available as well for Obs 5. The observations were performed with the ACIS in continuous-clocking (CC) mode. No gratings were employed. The X-ray data analysis was performed using the *Chandra* Interactive Analysis of Observation (CIAO) software 4.14.1 (Fruscione et al. 2006), with CALDB version 4.9.8.

We used a box of 20"×20" for source extraction, and computed responses with point-spread function corrections (see Appendix A and B for more details on the *Chandra* extractions and analyses). For background extraction, we used a source-free box of 20"×100". Data were analyzed from 1.4-10 keV.

## 3. ANALYSIS & RESULTS

Spectral analysis was carried out using XSPEC 12.13.0c (Arnaud 1996). The principal aim of our spectral analysis is to leverage the broadband X-ray data to examine the accretion geometry (and determine associated parameters of the BH system) throughout the stalled outburst. *Chandra* data are considered separately in Appendix A owing to complications related to photon pileup in the data. Briefly, our adopted self-consistent physical model is incompatible with use of the *pileup* model that is needed to correct *Chandra* spectral distortion. Unless specified otherwise, all uncertainties are presented at 90% confidence throughout.

### 3.1. *Preliminary Continuum Analysis*

As a first step, we fit the spectral continuum to characterize the data and search for the presence of reflection signatures like Fe-K$\alpha$ emission, as identified in MAXI J1813 in other works like Jiang et al. (2022) and Jana et al. (2021). We use *nthcomp* to model a Comptonization continuum spectrum (Życki et al. 1999), with neutral gas absorption modeled by *TBabs* (Wilms et al. 2000). A representative fit and data-to-model ratio for Obs 5 are shown in Figure 3. The residuals show broadened Fe-K fluorescence, indicative of relativistic reflection in the data. Our continuum analysis of all 7 observations produces results consistent with MAXI J1813 having remained in the canonical hard state for the duration of its outburst. The Compton power-law dominates the spectrum and exhibits a photon index $\Gamma=1.6–1.8$, typical of a hard state.

### 3.2. *Self-Consistent Spectral Model*

We next formulate a self-consistent spectral model which describes the Comptonization of seed thermal disk photons, and a reflection component associated with the illumination of the disk's surface by back-scattered Compton photons. Our model is constructed to require that *all* photons emerging from the disk's surface are Comptonized by the same corona, and so we impose the same Comptonization kernel on both direct disk emission and reflection emission components. Compared to models which neglect the role of Compton-scattering on the reflection emission, this self-consistent formalism generally increases the reflection fraction. The disk thermal component is modeled using the multicolor blackbody model *diskbb* (Makishima et al. 1986). This is defined by two parameters, a characteristic temperature $kT$, and a normalization. To model Comptonization, we use the convolution model *simplcut* which can be supplied with an arbitrary seed spectrum (in this case, the disk emission) (Steiner et al. 2009; Steiner et al. 2017; Steiner et al. 2017). Its key parameters are a photon index $\Gamma$, a scattering fraction $f_{sc}$ (i.e., the proportion of seed photons which are Comptonized by the corona), electron temperature $kT_e$, and the reflection fraction $R_f$ which describes the apportionment of Compton-scattered photons which illuminate the disk. Reflection is described using *relxillCp* version 2.1, from the *relxill* model library (Dauser et al. 2013; García & Kallman 2010). We employed simplcut in order to be able to also Comptonize the reflection component. Simplcut, $\Gamma$, $kT_e$ and $R_f$ parameters are matched in the two models, and both use nthcomp for the kernel, but simplcut is capable of working with an arbitrary seed spectral shape (here thermal emission and the reflection emission both). Moreover, by using simplcut in this way, we properly account for the fraction of coronal photons which hit the disk versus those which reach the observer; otherwise those two quantities aren't properly coupled. The radial reflection emissivity profile is defined by power-law indices $q_1$ and $q_2$ with a break radius between them; here we have simply set $q_1 = q_2 = 3$. The innermost disk radius $R_{in}$, BH spin $a_*$, and inclination $i$ together define the disk and spacetime geometry. We fix the spin to its maximum allowed value, $a_* = 0.998$, and fit for $R_{in}$ and $i$. The strength of Fe-K fluorescence and the Compton hump is modulated by the iron abundance, $Z_{Fe}$, which is a free parameter when fitting with *NuSTAR*. The disk's surface is approximated as a uniform slab of density $n$ with ionization parameter $\xi$.

The reflection component being described by *relxillCp* is produced when disk emission Compton scatters in the corona, and a fraction of those scattered, energized photons return to illuminate the surface of the disk. The emission seeding the coronal Compton signal is a combination of thermal emission produced by disk viscosity / reflection thermalization and reflection reprocessing at the surface of the disk. The present formulation of *relxill* and its variants don't allow for an arbitrary seed spectrum and instead is set by a power-law spectral illumination extending down to $\sim 0.1$ keV. Unfortunately, for disk spectra of interest for BHXRBs, this power-law approximation is problematic at low energies, especially those below the disk's characteristic temperature, as we illustrate in Figure 2, viz. compare the blue and green curves. Given that *simplcut* scatters photons between energies, it is important that the reflection continuum shape not behave unphysically out of the fitted bandpass, otherwise that unphysical behavior could alter the scattered signal over the observed bandpass. To correct for this unphysical runaway, we apply an empirical correction by defining a multiplicative factor that is unity at high energies but falls off as a power-law at low energies. This multiplicative broken power-law ("mbknpo") has two terms: the break energy, which has been fixed to $b = kT_{disk} \times 4.5$ from empirical exploration of Comptonization models, and the spectral index below the break which has a fixed relationship to the Compton power-law photon index: $I = \Gamma - 1.5$ (see Svoboda et al. 2024; Steiner et al. 2024). It's functionality is illustrated in the blue curve of Fig. 2.

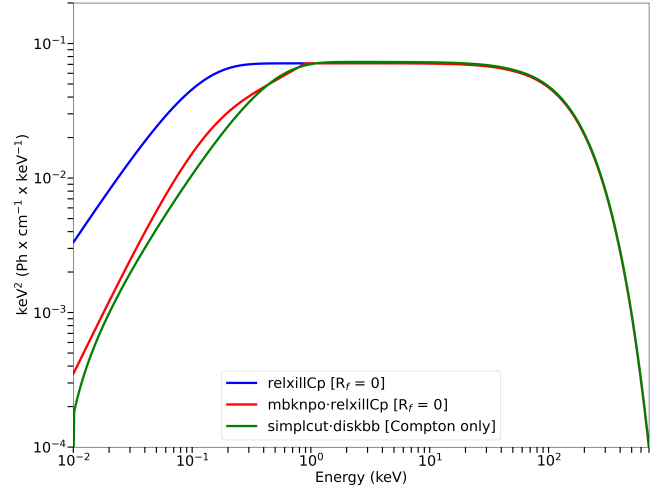

**Figure 2.** An illustration of the runaway flux problem when applying current reflection models to BHXRBs. The blue line shows the power-law spectrum adopted by *relxillCp* which illuminates the disk. The green line shows the desired ("true") spectrum corresponding to Compton-scattered emission from a 0.2 keV *diskbb* component. The red line shows our adopted multiplicative correction applied to the blue curve. This correction curtails the unphysical runaway at low energies and provides a reasonable approximation to the Comptonization spectrum.

Put together, our physical model is formulated as: $tbabs \times (simplcut(diskbb + mbknpo \times relxillCp))$. As a final addition, in order to account for cross-calibration differences between missions, which is often of order $\gtrsim 10\%$ in absolute flux, we include a flux standardization model *crabcorr* (Steiner et al. 2010). This multiplicative model was designed to standardize flux measurements of the Crab (or other calibration targets) between instruments. But where such measurements are variable in time, uncertain, or unavailable, its two parameters (a cross-normalization constant and a difference in spectral index, $\Delta\Gamma$) can also be fitted directly. In fullness, our spectral model becomes:
$tbabs \times (simplcut(diskbb + mbknpo \times relxillCp)) \times crabcorr$.

### 3.3. Fitting Results

We define a logarithmic energy grid from 10 eV to 500 keV in 500 intervals for computing the model, and apply this model to Obs 1–7. For Obs 1–5 which include *NICER* data, we adopt XSPEC's *pg*-statistic for fitting, which assumes Poisson-statistics for the data and Gaussian-statistics for the background uncertainty, as appropriate for *NICER* given use of the 3C50 background model. For obs 6 & 7, we instead use a $\chi^2$ fit statistic since these observations have no NICER data, and Gaussian statistics better capture the systematics present in the *Swift/Chandra/NuSTAR* data. We note that this choice of statistic has insignificant impact on the best-fitting parameter values in the resulting fit. We fix the neutral hydrogen column to $N_H = 1.35 \times 10^{22}$ cm$^{-2}$ based on our preliminary fits presented in Section 3.1. For Obs 1–4 which have no *NuSTAR* high-energy data, we fix $kT_e = 100$ keV and adopt a fixed iron abundance of $Z_{Fe} = 3$ based on fits to Obs 5–7.

We analyze data in the 0.7-10 keV energy range with *NICER* and *Swift*, an energy range of 3-75 keV with *NuSTAR*. In Table 2, we present the flux for each observation and the corresponding X-ray luminosity, assuming a fiducial distance of 8 kpc and mass of 10 $M_\odot$.

Our full fitting results to Obs 1–7 are presented in Table 3. We remark that, operationally, the reflection fraction $R_f$ is determined iteratively self-consistently by matching the disk illumination assumed by *relxillCp* to that produced by *simplcut* (see Steiner et al. 2016, 2017). In this way, the self-consistent $R_f$ is determined iteratively from the fit, parameterizing the relative strength of reflection flux observed compared to the coronal continuum flux observed (i.e., sometimes alternatvely termed as "reflection strength"). As shown in Table 3, the thermal component is consistent with a disk black-body emission of $kT_{in} \sim 100$-$300$ eV along the observations, which is consistent with the values given by Armas Padilla et al. (2019) and Jana et al. (2021) from joint *NuSTAR* and *Swift* observations corresponding to Obs. 5–7. According to these results, we conclude that any thermal emission must be cool and weak, as is typical for the hard state.

Our best-fitting results using *relxillCp* suggest a thin disk with an inner radius of $R_{in}$ 2.1+/−1.3$r_g$ around the BH (90% confidence). This value describes the weighted mean of the results in Table 3, a result which is strongly anchored in the high-precision fit obtained for Obs 5. Obs 5 is unique in having very high signal-to-noise across the full bandpass, as it is notably the *only* observation which includes both *NICER* and *NuSTAR* data. Because of the comparatively weaker constraining power of the other observations, we cannot rule out the possibility that Obs 5 alone has the inner-edge close in compared to the other observations, but we feel the most likely interpretation is that the disk inner-radius is stable and small throughout, and the fits are simply less constraining on $R_{in}$. Adopting this picture, the low weighted-mean of $R_{in}$ suggests that either the BH is very highly spinning and very slightly truncated, or that the disk extends to the ISCO and the BH spin is $a_* \gtrsim 0.7$. We find the inclination at values of $i = 14$-$47°$ throughout the observations with values in the last three observations closely matched by Jiang et al. (2022). We find that the inclination values are consistent with a constant value of $\sim 28° \pm 10°$ (90%), from the weighted mean. All the fits show a moderate-to-high ionization ($log\,\xi \sim 2-3.2$), with density ($n_e \sim 10^{16-19}$cm$^{-3}$).

Figure 4 presents spectral fits for two representative observations (Obs 1 and Obs 5) for our campaign in the topmost panels. The bottom series of panels depict data-to-model ra-tios, all tightly clustered around unity without obvious residual features. These illustrate the quality of our fits.

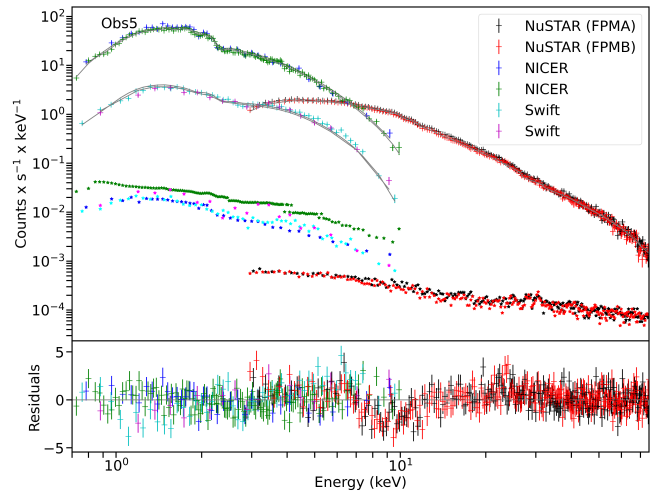

**Figure 3.** The spectrum for Obs 5. Top: *NICER* (blue and green), *Swift* (purple and cyan: XRT) and *NuSTAR* (black: FPMA; red: FPMB) source and associated background spectra (as stars) of MAXI J1813 for a continuum fit of *tbabs* × *nthcomp*. Bottom: Associated data-to-model ratio plot. The residuals – *NuSTAR* in particular – show a broad peak around 6.5 keV corresponding to Fe-K fluorescence, an indicator of spectral reflection. We applied a rebinning on *NICER* and *Swift* in order to make the Fe-K line more visible with *NuSTAR* showing a broad small peak. Two GTIs apiece for each of *Swift* and *NICER* are included in Observation 5. We fit for spectral differences between GTIs to account for potential variation.

**Table 2.** The observed flux of MAXI J1813. The X-ray luminosity is computed from the unabsorbed 0.7- 10 keV flux, adopting a distance of 8 kpc and assuming a BH mass of 10 solar masses in computing the Eddington luminosity.

| Observations | Flux$_{0.7-10\,(keV)}$ ($10^{-10}$erg·cm$^{-2}$·s$^{-1}$) | L$_x$/L$_{Edd}$ (unabs.) |
|---|---|---|
| Obs 1 | 8.6±0.1 | 0.008 |
| Obs 2 | 8.8±0.2 | 0.008 |
| Obs 3 | 8.1±0.2 | 0.007 |
| Obs 4 | 6.6±0.1 | 0.006 |
| Obs 5 | 6.2±0.2 | 0.004 |
| Obs 6 | 5.2±0.1 | 0.005 |
| Obs 7 | 5.8±0.1 | 0.004 |

Subsequent to optimizing our spectral fits, we determine the uncertainties using Markov-Chain Monte Carlo (MCMC). The XSPEC MCMC implementation is based on the work of Foreman-Mackey et al. (2013), who devised a affine-invariant sampler (Goodman & Weare 2010) consisting of a number of "walkers" which navigate and map out the posterior parameter space in proportion to its probability. Our MCMC runs consist of 54 walkers, with an initial chain length of 100000. After burning in using three iterations, we perform 10–50 million chain-length runs, having extended as needed to reach convergence which is essential for obtaining accurate estimates of parameters and credible intervals from MCMC simulations. The posterior uncertainties given in Table 3 come from this MCMC analysis, and we note that the ranges presented are consistent with results from using the "error" command in XSPEC. Figure 5 shows contour plots for Obs 1 and Obs 5, presenting pair-wise correlations between all combinations of $R_{\rm in}$, $log\xi$, $Z_\odot$, $kT_{\rm e}$ and $i$. On the diagonal, a histogram of each parameter can be seen. Notably, while $R_{\rm in}$ is only weakly constrained to be $\lesssim 80\,R_{\rm g}$ for Obs 1, in the case of Obs 5, $R_{\rm in}$ is very precisely constrained to $\approx 2.0 \pm 0.5\,R_g$ ($1\sigma$). We illustrate the performance of our MCMC runs in Figure 6, which depicts the autocorrelation of $R_{\rm in}$ and $i$ for Obs 5. As is shown here, we have run our chains to at least $\gtrsim 10$ times the typical autocorrelation length of a walker for the slowest-converging parameter in order to obtain good sampling of the posterior distribution. The MCMC results again confirm the key constraining power of observation 5, owing to the strong synergy between *NICER* and *NuSTAR* in particular.

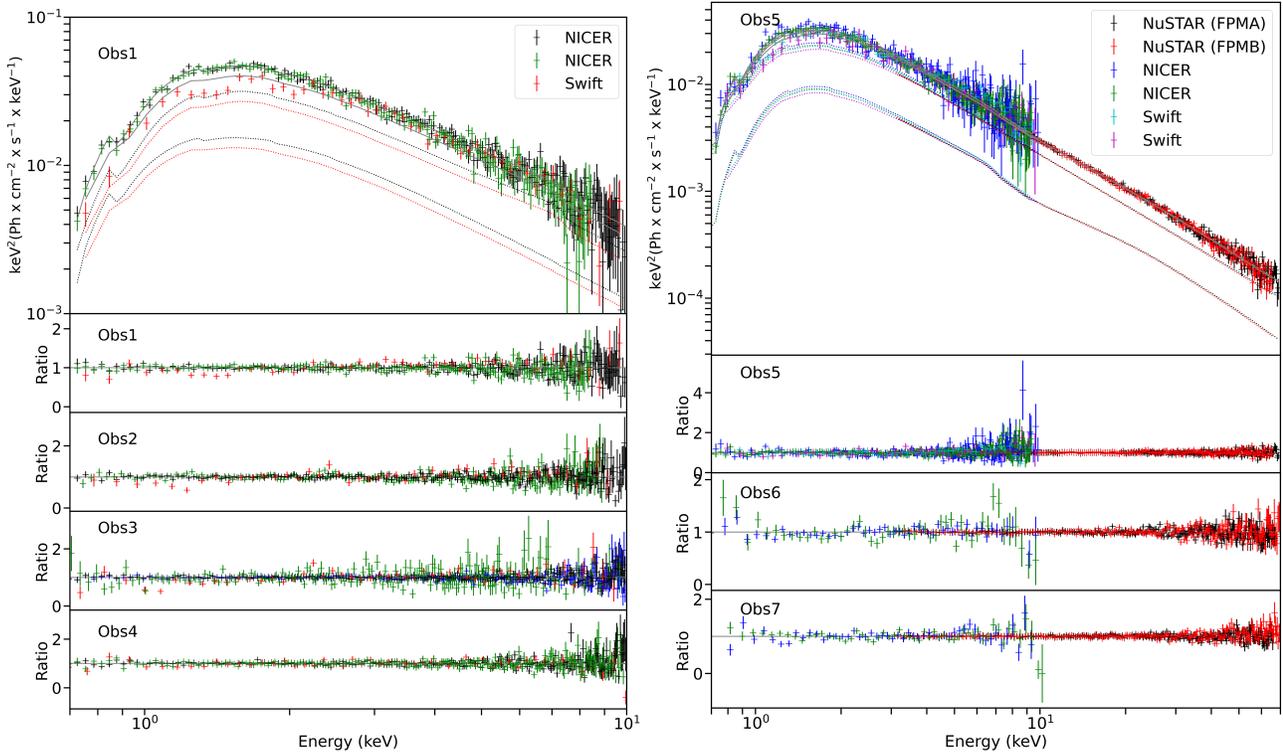

**Figure 4.** Top panels show the Obs 1 and Obs 5 unfolded spectra and the lower panels are the ratio plots of each observation with the best fit. Both spectra show the continuum and the reflection model. The two *NICER* GTIs shown with the best S/N ratio. For *Swift* we used only GTI due to only one was created for the left panel and two for the right panel.

## 4. DISCUSSION

We obtain the reflection fraction iteratively, by comparing the incident coronal emission which is assumed by *relxillCp* to reach the disk (operationally, obtained with a *relxillCp* component omitted from the fit for which $R_f$ is set to 0, and then scaled by $((1+R_f) \times R_f^{-1})$) with the Compton component being output from *simplcut* (Section 3.2), and adjusting them to match. The primary change from refitting while iteratively adjusting $R_f$ was on $f_{\rm sc}$. We required the incident and output Compton components to match within 5% in each fit.

Our previous investigation, Jiang et al. (2022), revealed evidence of reflected emission from the inner region of the disk. Here, in this follow-up work in which we impose a self-consistent modeling framework, we reach similar conclusions about several properties of the system, e.g., $R_{\rm in}$ and $i$. However, a crucial difference between our work and the analyses of Jiang et al. (2022) and Jana et al. (2021) is we find a much larger $R_f$, very close to unity. This difference is because here we account for the fact that a proportion of the disk reflection emission also scatters in the corona, resulting in a hardened continuum component which is mingled with the continuum produced by Compton-scattered thermal-disk emission. Consequently, when including this effect, the reflection fraction is several-fold larger than when this consideration is omitted. Moreover, this larger reflection fraction, of order unity, is consistent with most static coronal geometries (see, e.g., Dauser et al. 2016). The identifiable Fe-K$\alpha$ feature is associated with the transmitted portion of reflection emission passing through the corona. Other aspects of Jiang et al.

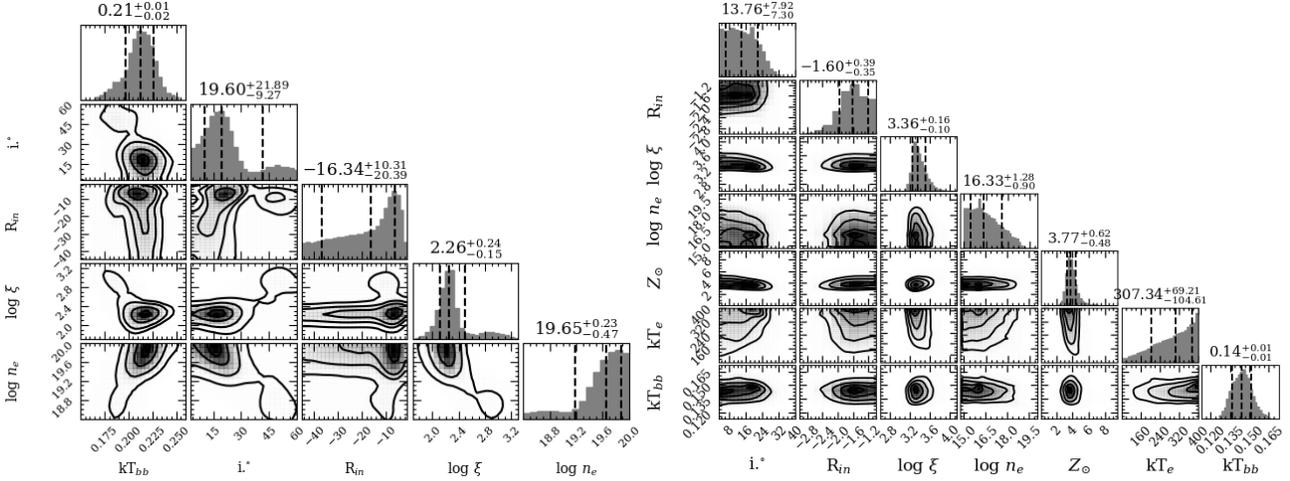

**Figure 5.** Corner plots show MCMC posterior distributions of parameters of interest for Obs 3 (left) and Obs 5 (right). We emphasize on the inner radius parameter showing a better constrain in Obs 5 although in both cases the 1$\sigma$ contour is showing a sligthly truncated or non-truncation of the disk. In both cases the disk is moderate to high ionized and dense.

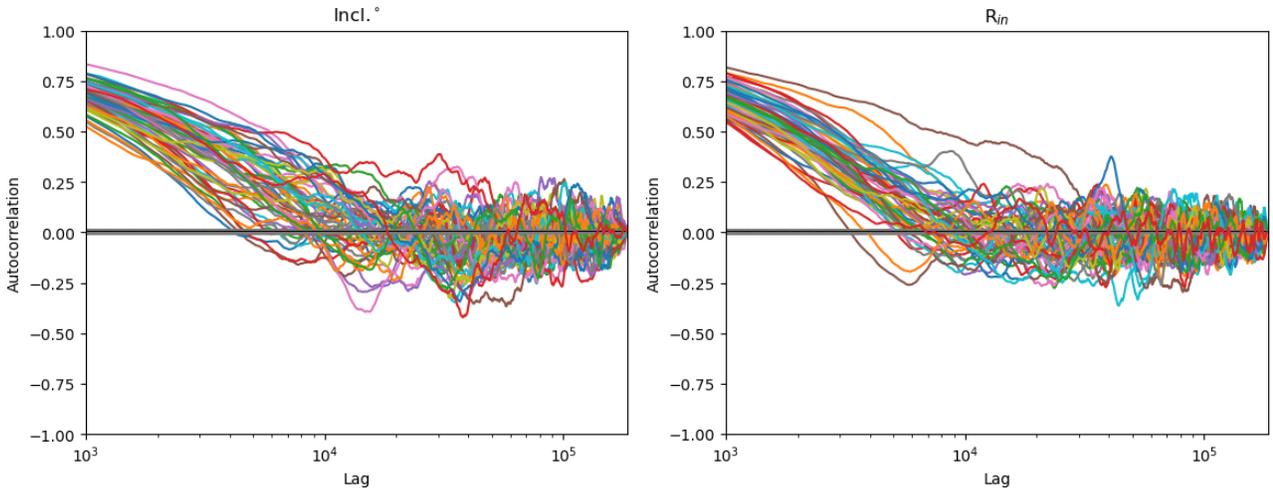

**Figure 6.** Autocorrelation lags for Obs 5 MCMC walkers for $i$ and $R_{\rm in}$. Each colored line shows the behavior of an individual walker. As shown here, the chains have been run to well over 10 times the characteristic autocorrelation length. This is important to ensure the converged chains obtain representative samples of the posterior parameter space with a useful effective sample size. Note how this measure can differ somewhat between different parameters.

(2022) are consistent with our analysis, including the disk ionization and the temperature of the cool disk emission. In contrast, Jana et al. (2021) fitted a significantly hotter thermal component. Jana et al. (2021) also fit the inner radius via the *laor* model obtaining an inner radius of $\sim 2.6 R_g$ quite similar to our constraint.

Our self-consistent disk reflection analysis with the *relxillcp* model, showed similarities with previous studies in the same epoch. The most important similarity is consistently small inner radius found in all three studies (all finding $R_{\rm in} < 7 R_g$) which rules against large-scale truncation for MAXI J1813's hard state peak. This is despite a putative low Ed-dington luminosity, for a nominal mass of $10 M_\odot$ and a fiducial distance of 8 kpc. This finding is noteworthy in that a topic of high interest over recent years in BHXRBs has been whether or not accretion disks are truncated at many tens or hundreds of $R_g$ in hard states. In particular there are disagreements between results based on timing data (e.g., Wang et al. 2021; De Marco et al. 2021), versus those determined using reflection spectroscopy (e.g., Jiang et al. 2019; García et al. 2019). Other spectral results have also generally favored interpretations in which at sufficiently high luminosities (above a few percent of $L_{\rm Edd}$), the disk is at the ISCO or else only slightly truncated (e.g. Reis et al. 2010; Plant et al. 2015).

**Table 3.** Best-fitting model parameters for Obs 1–7. The quoted errors are at the 90% confidence level based upon MCMC posteriors with uniform priors. For the purposes of fitting over the maximum allowed parameter space, the BH spin has been fixed at $a_* = 0.998$ with a free inner radius and $N_H$ fixed at $1.35 \times 10^{22}$ cm$^{-2}$ observations.

| Parameters | Obs 1 | Obs 2 | Obs 3 | Obs 4 | Obs 5 | Obs 6 | Obs 7 |
|---|---|---|---|---|---|---|---|
| $\Gamma$ | $1.69^{+0.04}_{-0.09}$ | $1.68^{+0.03}_{-0.14}$ | $1.65^{+0.03}_{-0.02}$ | $1.68^{+0.02}_{-0.03}$ | $1.61^{+0.01}_{-0.01}$ | $1.67^{+0.02}_{-0.02}$ | $1.79^{+0.05}_{-0.04}$ |
| $f_{sc}$ | $0.40^{+0.13}_{-0.24}$ | $0.39^{+0.13}_{-0.22}$ | $0.46^{+0.07}_{-0.11}$ | $0.34^{+0.12}_{-0.13}$ | $0.28^{+0.06}_{-0.06}$ | $0.41^{+0.25}_{-0.17}$ | $0.75^{+0.08}_{-0.21}$ |
| $R_f^*$ | $0.97^{+0.19}_{-0.25}$ | $0.75^{+0.26}_{-0.17}$ | $0.86^{+0.21}_{-0.06}$ | $0.67^{+0.1}_{-0.34}$ | $0.80^{+0.14}_{-0.02}$ | $0.76^{+0.13}_{-0.15}$ | $0.95^{+0.13}_{-0.11}$ |
| $kT_{disk}$ (keV) | $0.18^{+0.03}_{-0.03}$ | $0.18^{+0.36}_{-0.06}$ | $0.21^{+0.02}_{-0.03}$ | $0.16^{+0.06}_{-0.04}$ | $0.14^{+0.01}_{-0.01}$ | $0.12^{+0.05}_{-0.02}$ | $0.12^{+0.07}_{-0.02}$ |
| Norm$_{disk}$ | $(33^{+40}_{-20}) \times 10^3$ | $(15^{+80}_{-10}) \times 10^3$ | $(15^{+10}_{-10}) \times 10^3$ | $(42^{+20}_{-40}) \times 10^3$ | $(100^{+70}_{-40}) \times 10^3$ | $(16^{+20}_{-20}) \times 10^4$ | $(36^{+210}_{-30}) \times 10^3$ |
| Incl. (°) | $47^{+30}_{-40}$ | $43^{+20}_{-10}$ | $20^{+40}_{-10}$ | $37^{+30}_{-20}$ | $14^{+10}_{-10}$ | $40^{+10}_{-10}$ | $23^{+10}_{-10}$ |
| $R_{in}$ ($R_g$) | $34^{+77}_{-31}$ | $16^{+13}_{-13}$ | $20^{+36}_{-14}$ | $15^{+14}_{-13}$ | $2.0^{+0.8}_{-0.7}$ | $10^{+46}_{-9}$ | $12^{+36}_{-6}$ |
| log $\xi$ (erg cm s$^{-1}$) | $3.18^{+0.62}_{-1.20}$ | $2.81^{+0.66}_{-0.39}$ | $2.26^{+0.73}_{-0.26}$ | $2.46^{+0.73}_{-0.61}$ | $3.36^{+0.28}_{-0.16}$ | $1.87^{+0.66}_{-0.34}$ | $2.75^{+0.11}_{-0.97}$ |
| $Z_{Fe}(Z_\odot)^{**}$ | 3 | 3 | 3 | 3 | $3.8^{+1.2}_{-0.8}$ | $2.9^{+2.8}_{-1.7}$ | $1.1^{+1.1}_{-0.3}$ |
| log$n_e$ (cm$^{-3}$) | $17.6^{+1.7}_{-2.4}$ | $18.7^{+1.1}_{-2.6}$ | $19.6^{+0.3}_{-1.2}$ | $18.6^{+1.3}_{-2.6}$ | $16.3^{+2.0}_{-1.2}$ | $16.1^{+2.3}_{-0.9}$ | $16.1^{+1.3}_{-1}$ |
| $kT_e$ (keV)$^{***}$ | 100 | 100 | 100 | 100 | $307^{+90}_{-170}$ | $249^{+140}_{-170}$ | $294^{+100}_{-180}$ |
| Norm$_{Refl}$ | $(2.9^{+2.9}_{-1.6}) \times 10^{-3}$ | $(4.3^{+3.6}_{-1.4}) \times 10^{-3}$ | $(1.4^{+1.1}_{-0.5}) \times 10^{-2}$ | $(2.3^{+1.8}_{-0.9}) \times 10^{-3}$ | $(2.3^{+0.9}_{-0.9}) \times 10^{-3}$ | $(1.8^{+1.1}_{-0.5}) \times 10^{-3}$ | $(2.4^{+0.8}_{-0.8}) \times 10^{-3}$ |
| cross-norm (XRT) | $0.86^{+0.02}_{-0.02}$ | $0.94^{+0.03}_{-0.03}$ | $0.86^{+0.03}_{-0.02}$ | $1.07^{+0.03}_{-0.03}$ | $0.98^{+0.03}_{-0.03}$ | $0.96^{+0.02}_{-0.02}$ | $0.93^{+0.03}_{-0.02}$ |
| $\Delta\Gamma$ (XRT) | 0 | 0 | 0 | 0 | 0 | 0 | 0 |
| cross-norm. (FPMA) | - | - | - | - | 1 | 1 | 1 |
| $\Delta\Gamma$ (FPMA) | - | - | - | - | 0 | 0 | 0 |
| cross-norm. (FPMB) | - | - | - | - | $1.01^{+0.02}_{-0.02}$ | $1.00^{+0.02}_{-0.02}$ | $0.99^{+0.02}_{-0.01}$ |
| $\Delta\Gamma$ (FPMB) | - | - | - | - | $(1.0^{+0.7}_{-0.7}) \times 10^{-2}$ | $(1.6^{+0.7}_{-0.7}) \times 10^{-2}$ | $(6.4^{+7.5}_{-7.1}) \times 10^{-3}$ |
| cross-norm. (NICER) | 1 | 1 | 1 | 1 | $1.10^{+0.03}_{-0.03}$ | - | - |
| $\Delta\Gamma$ (NICER) | 0 | 0 | 0 | 0 | $(6.1^{+2.2}_{-2.3}) \times 10^{-2}$ | - | - |
| $\chi^2/\nu$ | (PG-stat)528/461 | (PG-stat)509/464 | (PG-stat)760/657 | (PG-stat)720/705 | (PG-stat)1056/928 | 585/522 | 570/536 |

* Reflection fraction values are determined iteratively by matching illuminating and scattered components self consistently. The errors bars have been assessed by repeating this process for random samples from the chain.
** The iron abundance is sensitive to the Compton hump strength and so has been fixed at 3 for Obs 1–4 without *NuSTAR* data based upon a round average from the fits to Obs 5–7.
*** The electron temperature is fixed to a round 100 keV for Obs 1–4 owing to a lack of high-energy spectral coverage in the absence of *NuSTAR* coverage.

The parameters of the reflection model across the 7 observations are heterogeneous in precision, but with consistent values. In particular, they are consistent with a single inner-disk / coronal geometry across the month-long span of the observations. The coronal temperature can only be constrained using high-energy *NuSTAR* data, and is consistently kT$_e$ > 100 keV for Obs 5–7, for which *NuSTAR* data is available. $R_f$ is consistent and near unity, ranging between ∼0.7–1.0. The data support a moderately-high density, $n_e \approx 10^{16} - 10^{19} cm^{-3}$, and a high metallicity, with $Z \approx 3Z_\odot$. As mentioned in Sec-tion 3.3, the strongest constraints are achieved from Obs 5, which is the only observation featuring broadband coverage with both *NuSTAR* and *NICER*, i.e., the two highest throughput facilities in their respective bandpasses. Because this constraint is so dominant over our net results, we have also investigated our joint constraints when omitting Obs 5. Without observation 5, in which we joint *NuSTAR*, *NICER* and *Swift*, the inner radius increases, at 90% confidence level to $\sim 16 \pm 0.7 R_g$ and the inclination goes up to $\sim 34 \pm 6°$. This increase in the inclination is still consistent with previous works and the inner radius appears to be slightly truncated.

Our parameters show a canonical hard state with typical values, showing a hot corona dominating the emission (> 100 keV), a thin cold disk ~0.2 keV and a spectral index of 1.6-1.8. The values for the disk ionization and disk density are moderate to high which is expected in such stage of an outburst. By imposing a self-consistent disk reflection modeling, all these parameters remain similar but we account for a Compton-scattered reflected photons, driving to a higher reflection fraction close to unity. These reflection values close to unity could be indicative of a compact corona or a corona closer to the compact object. Contrary to the typical picture of a hard state, the inner disk is close to the ISCO.

We turn to two recurrent black hole transients which are useful touchstones with which to contextualize our results: GX 339-4 and H1743-322. Both systems have presented full outbursts and also recent stalled outbursts, which we use as a point of comparison here. In 2018, H1743-322 had a month-long stalled outburst observed by *Swift*, *NICER*, *NuSTAR* and *XMM-Newton*, which was analyzed by Stiele & Kong (2021). It's X-ray lightcurve and hardness evolution is very similar that exhibited by MAXI J1813. It's peak brightness in this stalled outburst was in the vicinity 10-20 mCrab, whereas it's historical brightness peak has been ≳ 1 Crab (Remillard & McClintock 2006). Stiele & Kong (2021) identify a similar $\Gamma \approx 1.65$ and a similarly cool accretion disk component $kT_{\rm disk} \approx 0.2$ keV. The disk ionization measured is relatively low, $\log \xi \approx 1.7 - 1.8$, and best estimate for the disk inner radius found to be $\sim 6 - 8 \ R_g$. This large $R_{\rm in}$ and low ionization broadly comport with the expectation of its likely having low spin (Steiner et al. 2012; Nathan et al. 2024), and would be consistent with moderate-to-no truncation. Notably, this analysis was analogous to our earlier non-self-consistent modeling of MAXI J1813 (Jiang et al. 2022), and just as the $R_f$ was low in our earlier analysis, for H1743-322 the *relxill* fits returned a low reflection fraction ($\sim 0.3$). As the reflection fraction increased $\sim$ 5-fold in this analysis from imposing self-consistency with the Comptonization of reflection emission, we expect this model would return a similarly larger $R_f$ (closer to unity) for those H1743-322 data (see Dauser et al. 2016).

However, the 2017 stalled outburst of GX 339-4 presents very differently. That outburst, which lasted $\sim$ 5 months, was examined in García et al. (2019), primarily focused on two epochs using *Swift* and *NuSTAR* data. As for H1743-322, these observations were at a flux several percent of the historical peak of GX 339-4. The best-fitting model adopted in that analysis utilized a dual-lamppost coronal model for which the spectrum was canonically hard $\Gamma \sim 1.5$. The closer-in reflection component required very high ionization $\log \xi \sim 4$. Their fits found modest-to-no disk truncation and require that the spin must be quite high. This analysis also did not impose the self-consistent coronal scattering of reflection emission, but notably the reflection *strength* in their model was already very high; $\sim$ 2. Applying the self-consistent model would result in a higher-still reflection strength for that adopted picture. Alternatively, we tentatively suggest that the near-featureless reflection continuum associated with $\log \xi \sim$ 4 would naturally be described by the Comptonized reflection emission from a *single* coronal component, negating the need for the more complex model adopted there. Such an analysis, however, is outside the scope of this work.

We also explored the fits to Obs 5 in search of statistically meaningful departure of the inner emissivity from the nominal profile $q = 3$. This was accomplished by freeing the inner emissivity ($q_1$) while fitting for a break radius, taken to be near $\sim 10 R_g$. However, the improvement in the fit was insignificant and more importantly, with the marginal improvement associated with a moderately higher q1, the inner radius became slightly smaller (while still falling within uncertainties). We also used the constraint on spin to very approximately estimate a possible distance of the source, making use of the thermal disk component. Here, we exchange *diskbb* for the relativistic disk model *kerrbb* (Ross & Fabian 2005). We adopt the same fiducial mass, 10 $M_\odot$, a spin and inclination linked to the reflection fits to Obs 5, and adopt a nominal color correction factor of 1.7. We fit for just the mass accretion rate and the distance, with a best-fitting distance of ~1.5 kpc and a corresponding $L/L_{\rm Edd} \lesssim 0.1\%$. We have also explored a fit to Obs 5 in which, rather than fitting for radius, we assume the disk reaches the ISCO and fit instead for the spin. This results in a constraint that spin $a_* > 0.77$ and a best fit of $a_* \sim 0.9$, which is consistent with our benchmark fits using a free $R_{\rm in}$.

## 5. CONCLUSIONS

MAXI J1813-095 underwent a stalled outburst in 2018 during which it remained in a canonical hard state with $\Gamma$ = 1.6-1.8, while exhibiting substantial spectral reflection. Assuming fiducial BH mass and distances of 10 $M_\odot$ BH and 8 kpc, respectively, the peak luminosity of this outburst is estimated as ≲1% of the Eddington luminosity. We present a self-consistent spectral analysis of the Comptonized reflection and disk continuum components, from which we constrain the inclination and inner radius to $28 \pm 6.5°$ and $\sim 2.1 \pm 0.8$ $R_g$, respectively. Conservatively adopting an extreme distance ($\sim 30$ kpc), we can confidently rule out large-scale truncation for L $\lesssim$ 10% Eddington luminosity for this stalled outburst.

This is consistent with either a moderate spin ($a_* \gtrsim 0.7$) and no disk truncation, or else mild disk truncation for an extreme-spin BH ($a_* > 0.9$). By having imposed self-consistency in the model by scattering not just the thermal disk emission, but also reflection emission, in the corona, we find that the reflection fraction is much larger compared to earlier analyses omitting this consideration. Our higher reflection fraction, $R_f \approx 0.7 - 1.0$ is in line with expectation for typical static-coronal geometry, and so supports the viability of a static - rather than outflowing - corona. We find that for MAXI J1813, self-consistency of the coronal scattering does not substantially impact conclusions regarding the BH spin or inclination.

## ACKNOWLEDGMENTS

This work was supported by Chandra General Observer grants GO8-19038A and GO9-20041X.

The authors thank Keith Arnaud for helpful discussions regarding XSPEC, and the nuances of the "energies" command and the "pileup" model. This study was conducted during the pandemic worldwide COVID-19 in 2021-2022. This research used data products and additional software tools provided by the High-Energy Astrophysics Science Archive Research Center (HEASARC). S.U. thanks the *NICER*, *NuSTAR*, *Swift*, and *Chandra* teams for their work on these observations and responsiveness to our queries. J.J. acknowledges support from Leverhulme Trust, Isaac Newton Trust and St. Edmund's College.

This research has made use of data from the NuSTAR mission, a project led by the California Institute of Technology, managed by the Jet Propulsion Laboratory, and funded by the National Aeronautics and Space Administration. Data analysis was performed using the NuSTAR Data Analysis Software (NuSTARDAS), jointly developed by the ASI Science Data Center (SSDC, Italy) and the California Institute of Technology (USA).

## DATA AVAILABILITY

This paper makes use of *NICER*, *Swift*/XRT, Chandra and *NuSTAR* archival data which can be acquired from HEASARC: https://heasarc.gsfc.nasa.gov/cgi-bin/W3Browse/w3browse.pl. This paper employs a list of Chandra datasets, obtained by the Chandra X-ray Observatory, contained in DOI: 20264, 20265, 20266.

## REFERENCES

Ajith, P., Hannam, M., Husa, S., et al. 2011, Phys. Rev. Lett., 106, 241101, doi: 10.1103/PhysRevLett.106.241101

Alabarta, K., Altamirano, D., Méndez, M., et al. 2021, MNRAS, 507, 5507, doi: 10.1093/mnras/stab2241

Armas Padilla, M., Muñoz-Darias, T., Sánchez-Sierras, J., et al. 2019, MNRAS, 485, 5235, doi: 10.1093/mnras/stz737

Armas Padilla, M., Muñoz-Darias, T., Sánchez-Sierras, J., et al. 2019, Monthly Notices of the Royal Astronomical Society, 485, 5235, doi: 10.1093/mnras/stz737

Arnaud, K. A. 1996, in Astronomical Society of the Pacific Conference Series, Vol. 101, Astronomical Data Analysis Software and Systems V, ed. G. H. Jacoby & J. Barnes, 17

Bakala, P., Török, G., Karas, V., et al. 2014, Monthly Notices of the Royal Astronomical Society, 439, 1933, doi: 10.1093/mnras/stu076

Bambi, C. 2018, Annalen der Physik, 530, doi: 10.1002/andp.201700430

Belloni, T. M. 2010, in Lecture Notes in Physics, Berlin Springer Verlag, ed. T. Belloni, Vol. 794, 53, doi: 10.1007/978-3-540-76937-8_3

Burrows, D. N., Hill, J. E., Nousek, J. A., et al. 2005, Space Science Reviews, 120, 165, doi: 10.1007/s11214-005-5097-2

Chruściel, P. T., Costa, J. L., & Heusler, M. 2012, Living Reviews in Relativity, 15, doi: 10.12942/lrr-2012-7

Connors, P. A., Piran, T., & Stark, R. F. 1980, ApJ, 235, 224, doi: 10.1086/157627

Cui, W., Ebisawa, K., Dotani, T., & Kubota, A. 1998, The Astrophysical Journal, 493, L75, doi: 10.1086/311134

Dauser, T., García, J., Walton, D. J., et al. 2016, A&A, 590, A76, doi: 10.1051/0004-6361/201628135

Dauser, T., Garcia, J., Wilms, J., et al. 2013, MNRAS, 430, 1694, doi: 10.1093/mnras/sts710

Davis, J. E. 2001, ApJ, 562, 575, doi: 10.1086/323488

De Marco, B., Zdziarski, A. A., Ponti, G., et al. 2021, A&A, 654, A14, doi: 10.1051/0004-6361/202140567

Debnath, D., Jana, A., Chakrabarti, S. K., Chatterjee, D., & Mondal, S. 2017, The Astrophysical Journal, 850, 92, doi: 10.3847/1538-4357/aa9077

Dovčiak, M., Muleri, F., Goosmann, R. W., Karas, V., & Matt, G. 2008, Monthly Notices of the Royal Astronomical Society, 391, 32, doi: 10.1111/j.1365-2966.2008.13872.x

Dunn, R. J. H., Fender, R. P., Körding, E. G., Belloni, T., & Cabanac, C. 2010, Monthly Notices of the Royal Astronomical Society, 403, 61, doi: 10.1111/j.1365-2966.2010.16114.x

Esin, A. A., McClintock, J. E., & Narayan, R. 1997, The Astrophysical Journal, 489, 865, doi: 10.1086/304829

Fabian, A. C., Rees, M. J., Stella, L., & White, N. E. 1989, Monthly Notices of the Royal Astronomical Society, 238, 729, doi: 10.1093/mnras/238.3.729

Fender, R. P. 2001, MNRAS, 322, 31, doi: 10.1046/j.1365-8711.2001.04080.x

Fender, R. P., Belloni, T. M., & Gallo, E. 2004, Monthly Notices of the Royal Astronomical Society, 355, 1105, doi: 10.1111/j.1365-2966.2004.08384.x

Foreman-Mackey, D., Hogg, D. W., Lang, D., & Goodman, J. 2013, Publications of the Astronomical Society of the Pacific, 125, 306, doi: 10.1086/670067

Fruscione, A., McDowell, J. C., Allen, G. E., et al. 2006, in Society of Photo-Optical Instrumentation Engineers (SPIE) Conference Series, Vol. 6270, Society of Photo-Optical Instrumentation Engineers (SPIE) Conference Series, ed. D. R. Silva & R. E. Doxsey, 62701V, doi: 10.1117/12.671760

Fuerst, F., Belanger, G., Parker, M., et al. 2018, The Astronomer's Telegram, 11357, 1

García, J., & Kallman, T. R. 2010, ApJ, 718, 695, doi: 10.1088/0004-637X/718/2/695

García, J. A., Steiner, J. F., McClintock, J. E., et al. 2015, The Astrophysical Journal, 813, 84, doi: 10.1088/0004-637x/813/2/84

García, J. A., Tomsick, J. A., Sridhar, N., et al. 2019, ApJ, 885, 48, doi: 10.3847/1538-4357/ab384f

García, J. A., Tomsick, J. A., Sridhar, N., et al. 2019, The Astrophysical Journal, 885, 48, doi: 10.3847/1538-4357/ab384f

Gendreau, K. C., Arzoumanian, Z., & Okajima, T. 2012, in Society of Photo-Optical Instrumentation Engineers (SPIE) Conference Series, Vol. 8443, Space Telescopes and Instrumentation 2012: Ultraviolet to Gamma Ray, ed. T. Takahashi, S. S. Murray, & J.-W. A. den Herder, 844313, doi: 10.1117/12.926396

Germanà, C. 2017, Phys. Rev. D, 96, 103015, doi: 10.1103/PhysRevD.96.103015

Goodman, J., & Weare, J. 2010, Communications in Applied Mathematics and Computational Science, 5, 65, doi: 10.2140/camcos.2010.5.65

Harrison, F. A., Craig, W. W., Christensen, F. E., et al. 2013, The Astrophysical Journal, 770, 103, doi: 10.1088/0004-637x/770/2/103

Hertz, P., & Grindlay, J. E. 1983, ApJ, 275, 105, doi: 10.1086/161516

Homan, J., & Belloni, T. 2005, Ap&SS, 300, 107, doi: 10.1007/s10509-005-1197-4

Homan, J., Miller, J. M., Wijnands, R., et al. 2005, The Astrophysical Journal, 623, 383, doi: 10.1086/424994

Jana, A., Jaisawal, G. K., Naik, S., et al. 2021, Research in Astronomy and Astrophysics, 21, 125, doi: 10.1088/1674-4527/21/5/125

Jiang, J., Fabian, A. C., Wang, J., et al. 2019, Monthly Notices of the Royal Astronomical Society, 484, 1972, doi: 10.1093/mnras/stz095

Jiang, J., Buisson, D. J. K., Dauser, T., et al. 2022, MNRAS, 514, 1952, doi: 10.1093/mnras/stac1401

Jordán, A., Côté, P., Ferrarese, L., et al. 2004, ApJ, 613, 279, doi: 10.1086/422545

Kalemci, E., Kara, E., & Tomsick, J. A. 2022, in Handbook of X-ray and Gamma-ray Astrophysics, 9, doi: 10.1007/978-981-16-4544-0_100-1

Karssen, G. D., Bursa, M., Eckart, A., et al. 2017, MNRAS, 472, 4422, doi: 10.1093/mnras/stx2312

Kawase, T., Negoro, H., Yoneyama, T., et al. 2018, The Astronomer's Telegram, 11323, 1

Kennea, J. A., Palmer, D. M., Lien, A. Y., et al. 2018, The Astronomer's Telegram, 11326, 1

Kostić, U., Čadež, A., Calvani, M., & Gomboc, A. 2009, AA, 496, 307, doi: 10.1051/0004-6361/200811059

Krawczynski, H., Muleri, F., Dovčiak, M., et al. 2022, Science, 378, 650, doi: 10.1126/science.add5399

Li, L.-X., Narayan, R., & McClintock, J. E. 2009, The Astrophysical Journal, 691, 847, doi: 10.1088/0004-637X/691/1/847

Liu, Q. Z., van Paradijs, J., & van den Heuvel, E. P. J. 2006, A&A, 455, 1165, doi: 10.1051/0004-6361:20064987

—. 2007, A&A, 469, 807, doi: 10.1051/0004-6361:20077303

Makishima, K., Maejima, Y., Mitsuda, K., et al. 1986, ApJ, 308, 635, doi: 10.1086/164534

McClintock, J. E., Narayan, R., & Steiner, J. F. 2014, SSRv, 183, 295, doi: 10.1007/s11214-013-0003-9

Nathan, E., Ingram, A., Steiner, J. F., et al. 2024, MNRAS, doi: 10.1093/mnras/stae1896

Plant, D. S., Fender, R. P., Ponti, G., Muñoz-Darias, T., & Coriat, M. 2015, A&A, 573, A120, doi: 10.1051/0004-6361/201423925

Plucinsky, P. P., Bogdan, A., Marshall, H. L., & Tice, N. W. 2018, in Space Telescopes and Instrumentation 2018: Ultraviolet to Gamma Ray, ed. J.-W. A. den Herder, K. Nakazawa, & S. Nikzad (SPIE), doi: 10.1117/12.2312748

Rao, A. R., Yadav, J. S., & Paul, B. 2000, The Astrophysical Journal, 544, 443, doi: 10.1086/317168

Rau, A. 2018, The Astronomer's Telegram, 11332, 1

Reis, R. C., Fabian, A. C., & Miller, J. M. 2010, Monthly Notices of the Royal Astronomical Society, 402, 836, doi: 10.1111/j.1365-2966.2009.15976.x

Reis, R. C., Miller, J. M., Reynolds, M. T., et al. 2013, The Astrophysical Journal, 763, 48, doi: 10.1088/0004-637x/763/1/48

Remillard, R. A., & McClintock, J. E. 2006, ARA&A, 44, 49, doi: 10.1146/annurev.astro.44.051905.092532

Remillard, R. A., Loewenstein, M., Steiner, J. F., et al. 2022, The Astronomical Journal, 163, 130, doi: 10.3847/1538-3881/ac4ae6

Reynolds, C. S. 2021, Annual Review of Astronomy and Astrophysics, 59, 117–154, doi: 10.1146/annurev-astro-112420-035022

Ross, R. R., & Fabian, A. C. 2005, MNRAS, 358, 211, doi: 10.1111/j.1365-2966.2005.08797.x

Russell, T. D., Miller-Jones, J. C. A., Sivakoff, G. R., Tetarenko, A. J., & JACPOT XRB Collaboration. 2018, The Astronomer's Telegram, 11356, 1

Santamaría, L., Ohme, F., Ajith, P., et al. 2010, Phys. Rev. D, 82, 064016, doi: 10.1103/PhysRevD.82.064016

Shidatsu, M., Ueda, Y., Nakahira, S., et al. 2011, Publications of the Astronomical Society of Japan, 63, S803, doi: 10.1093/pasj/63.sp3.S803

Steiner, J. F., García, J. A., Eikmann, W., et al. 2017, ApJ, 836, 119, doi: 10.3847/1538-4357/836/1/119

Steiner, J. F., García, J. A., Eikmann, W., et al. 2017, The Astrophysical Journal, 836, 119, doi: 10.3847/1538-4357/836/1/119

Steiner, J. F., McClintock, J. E., & Reid, M. J. 2012, ApJL, 745, L7, doi: 10.1088/2041-8205/745/1/L7

Steiner, J. F., McClintock, J. E., Remillard, R. A., et al. 2010, ApJL, 718, L117, doi: 10.1088/2041-8205/718/2/L117

Steiner, J. F., Narayan, R., McClintock, J. E., & Ebisawa, K. 2009, PASP, 121, 1279, doi: 10.1086/648535

Steiner, J. F., Remillard, R. A., García, J. A., & McClintock, J. E. 2016, ApJL, 829, L22, doi: 10.3847/2041-8205/829/2/L22

Steiner, J. F., Nathan, E., Hu, K., et al. 2024, arXiv e-prints, arXiv:2406.12014, doi: 10.48550/arXiv.2406.12014

Stella, L., & Vietri, M. 1998, ApJL, 492, L59, doi: 10.1086/311075

Stella, L., & Vietri, M. 1999, Phys. Rev. Lett., 82, 17, doi: 10.1103/PhysRevLett.82.17

Stiele, H., & Kong, A. K. H. 2021, ApJ, 914, 93, doi: 10.3847/1538-4357/abfaa5

Svoboda, J., Dovčiak, M., Steiner, J. F., et al. 2024, ApJL, 966, L35, doi: 10.3847/2041-8213/ad402e

Čadež, A., Calvani, M., & Kostić, U. 2008, A&A, 487, 527, doi: 10.1051/0004-6361:200809483

Vitale, S., Lynch, R., Raymond, V., et al. 2017, Phys. Rev. D, 95, 064053, doi: 10.1103/PhysRevD.95.064053

Wang, J., Mastroserio, G., Kara, E., et al. 2021, The Astrophysical Journal Letters, 910, L3, doi: 10.3847/2041-8213/abec79

Weisskopf, M. C., Tananbaum, H. D., Van Speybroeck, L. P., & O'Dell, S. L. 2000, in Society of Photo-Optical Instrumentation Engineers (SPIE) Conference Series, Vol. 4012, X-Ray Optics, Instruments, and Missions III, ed. J. E. Truemper & B. Aschenbach, 2–16, doi: 10.1117/12.391545

Wilms, J., Allen, A., & McCray, R. 2000, ApJ, 542, 914, doi: 10.1086/317016

Yu, W., & Yan, Z. 2009, The Astrophysical Journal, 701, 1940, doi: 10.1088/0004-637x/701/2/1940

Zhang, S. N., Cui, W., & Chen, W. 1997, The Astrophysical Journal, 482, L155, doi: 10.1086/310705

Życki, P. T., Done, C., & Smith, D. A. 1999, MNRAS, 309, 561, doi: 10.1046/j.1365-8711.1999.02885.x

APPENDIX

## A. PILE-UP IN ACIS CC-MODE DATA

Photon pileup arises due to a high flux of X-ray photons reaching the detector. When two different X-ray photons strike the same or neighboring pixels within the detector's integration time, their signals are conflated as a single event. This results in a phenomenon known as "pileup", where the signals from multiple photons become merged and/or the event grade (sometimes called the "pattern") can be distorted, impacting spectral analysis by rejecting some real events and combining energies for others. *Chandra* observed this source on 02-28-2018, 03-06-2018 and 03-25-2018 in continuous clocking-mode (CC-mode), with data showing signs of photon pileup. To assess this, we compared our spectra without the central pixel, for which pileup is worst, to a spectrum including the central pixel and another spectrum excluding central pixel, one pixel up and one pixel down. In Figure 7 we show the three ratio plots to the spectra previously mentioned, in which we clearly see photon pileup manifest from 6-10 keV. To minimize these effects we obtained two more spectra excluding the central pixel and the two nearest pixels to the central pixel. We found that when excluding the central pixel the pileup effect between 6-10 keV decreases substantially. For this reason, we find it necessary to apply a corrective pile-up model and next fit *Chandra* with the other instruments and confirm that *Chandra* data are in agreement with the analysis performed in Section 3.

We employ the *pileup* model (Davis 2001) to correct for the redistribution of soft X-ray photons events in energy and grade. The *pileup* model was designed for correction for *Chandra* imaging data and operates by simulating the effects of photon pileup on the observed spectrum. It convolves an input spectral model with a pileup response. Notably, CC-mode data is 1-dimensional and so the pileup effects are likely to be somewhat distorted compared to 2-dimensional data. We accordingly allow for the frametime parameter — which for imaging data would usually be fixed – to be free. In our analysis, data below 1.4 keV were ignored owing to the contamination layer on the ACIS filters (see Plucinsky et al. (2018)).

## B. CHANDRA ANALYSIS

The *pileup* model is unfortunately incompatible with *simplcut*. The former requires the default energy grid to be used for computations, whereas the latter requires an extended energy grid to compute redistribution properly. As a consequence, we adopt an approximate model which describes the same behavior as in our main analysis, but here we substitute *simplcut* with *nthcomp* as in Jiang et al. (2022).

In Table 4, we present the best fit including *Chandra* data for observations 5, 6 and 7. We set the same free parameters as those in the standard analysis and add the *pileup* model. The *Chandra pileup* model fits for the frametime and the alpha parameter, which $\alpha$ parameterizes "grade migration" in the detector, and represents the probability, per photon count greater than one, that the piled event is not rejected by the spacecraft software as a "bad event". Specifically, if N photons are piled together in a single frame, the probability of them being retained (as a single photon event with their summed energy) is given by $\alpha^{(N-1)}$. The true frametime of CC-mode is 2.85ms, but the best fit is a factor ∼3 larger. We speculate that this is a reasonable increase given the compression from a 2-dimensional image to 1-dimensional streak.

As a bottom line, we obtain broadly similar physical quantities as reflected in Table 3. Including *Chandra* data we didn't fit it self-consistently so the $R_f$ is unreliable compared to the Table 3. Figure 8 shows the three spectra with Chandra that corroborate that Chandra observations are consistent with the other instrument observations. We find that this practical approach, namely fitting CC-mode data with the *pileup* model using a fitted frametime, is successful at reconciling the mildly piled-up *CC-mode* data with the other observatories.

**Table 4.** Best-fit parameters for the averaged *NICER* (Ni), *Swift* (S), *NuSTAR* (Nu) and *Chandra* (Ch) spectra of MAXI J1813. The quoted errors are at the 90% confidence level with a BH spin fixed at 0.998, $N_H$ fixed at $1.35\times10^{22}$ cm$^{-2}$ and reflection fraction fixed at 1 along the observations. Obsid5-Obsid7

| Parameters | Obs 5 (Ni5+S5+Nu1+Ch1) | Obs 6 (S6+Nu2+Ch2) | Obs 7 (S7+Nu3+Ch3) |
|---|---|---|---|
| $\Gamma$ | $1.54^{+0.01}_{-0.01}$ | $1.66^{+0.04}_{-0.02}$ | $1.69^{+0.01}_{-0.01}$ |
| $kT_{disk}$ | $0.14^{+0.01}_{-0.01}$ | $0.51^{+0.03}_{-0.4}$ | $0.46^{+0.01}_{-0.01}$ |
| Norm.$_{disk}$ | $(6^{+0.6}_{-0.6})\times10^{-2}$ | $(8^{+3}_{-0.3})\times10^{-2}$ | $(9^{+0.4}_{-0.4})\times10^{-2}$ |
| Incl. | $12^{+12}_{-7}$ | $42^{+14}_{-4}$ | $25^{+5}_{-7}$ |
| $R_{in}$ ($R_g$) | $8.2^{+2.8}_{-1.6}$ | $8.8^{+12.3}_{-5.9}$ | $7.7^{+3.2}_{-2.4}$ |
| log $\xi$ (erg cm s$^{-1}$) | $3.65^{+0.04}_{-0.04}$ | $1.99^{+0.11}_{-0.25}$ | $3.1^{+0.1}_{-0.1}$ |
| $Z_{Fe}(Z_\odot)$ | $0.8^{+0.1}_{-0.1}$ | $0.98^{+1}_{-0.3}$ | $1.8^{+0.7}_{-0.5}$ |
| log $n$(cm$^{-3}$) | $19.8^{+0.1}_{-0.4}$ | 20 | $16^{+1.5}_{-0.8}$ |
| $kT_e$(keV) | $254^{+109}_{-90}$ | >267 | $189^{+160}_{-110}$ |
| Norm.$_{Refl}$ | $(1.5^{+0.2}_{-0.2})\times10^{-3}$ | $(1^{+0.7}_{-0.2})\times10^{-3}$ | $(7^{+1}_{-1})\times10^{-4}$ |
| Frame time | $(5.3^{+3.2}_{-1.6})\times10^{-3}$ | $(8.8^{+8.7}_{-4.1})\times10^{-3}$ | $(7^{+5}_{-3})\times10^{-3}$ |
| Alpha | $0.67^{+0.26}_{-0.22}$ | $0.47^{+0.35}_{-0.14}$ | $0.5^{+0.4}_{-0.2}$ |
| dNorm. (ACIS) | $0.85^{+0.03}_{-0.02}$ | 1.51 | 1.58 |
| d$\Gamma$ (ACIS) | 0 | 0 | 0 |
| dNorm.** (XRT) | $0.91^{+0.02}_{-0.02}$ | $0.90^{+0.01}_{-0.01}$ | 0.85 |
| d$\Gamma$ (XRT) | 0 | 0 | 0 |
| dNorm. (FPMA) | 1 | 1 | 1 |
| d$\Gamma$ (FPMA) | 0 | 0 | 0 |
| dNorm. (FPMB) | $0.99^{+0.003}_{-0.003}$ | $0.97^{+0.004}_{-0.004}$ | 0.98 |
| d$\Gamma$ (FPMB) | 0 | 0 | 0 |
| dNorm. (NICER) | $0.96^{+0.02}_{-0.02}$ | - | - |
| d$\Gamma$ (NICER) | $(8.6^{+1.2}_{-1.2})\times10^{-2}$ | - | - |
| $\chi^2/\nu$ | 1289/1077 | 1040/651 | 881/653 |

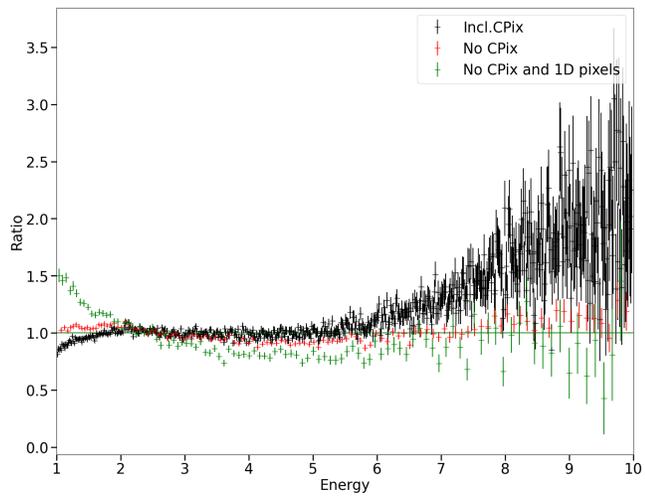

**Figure 7.** In this figure shows three ratio Chandra spectres plot (black full source, red excluding the central pixel and in green excluding the central pixel and the immediately adjacent pixels up and down) for observation 5 with a simple model *TBabs(simplcut(diskbb))*.

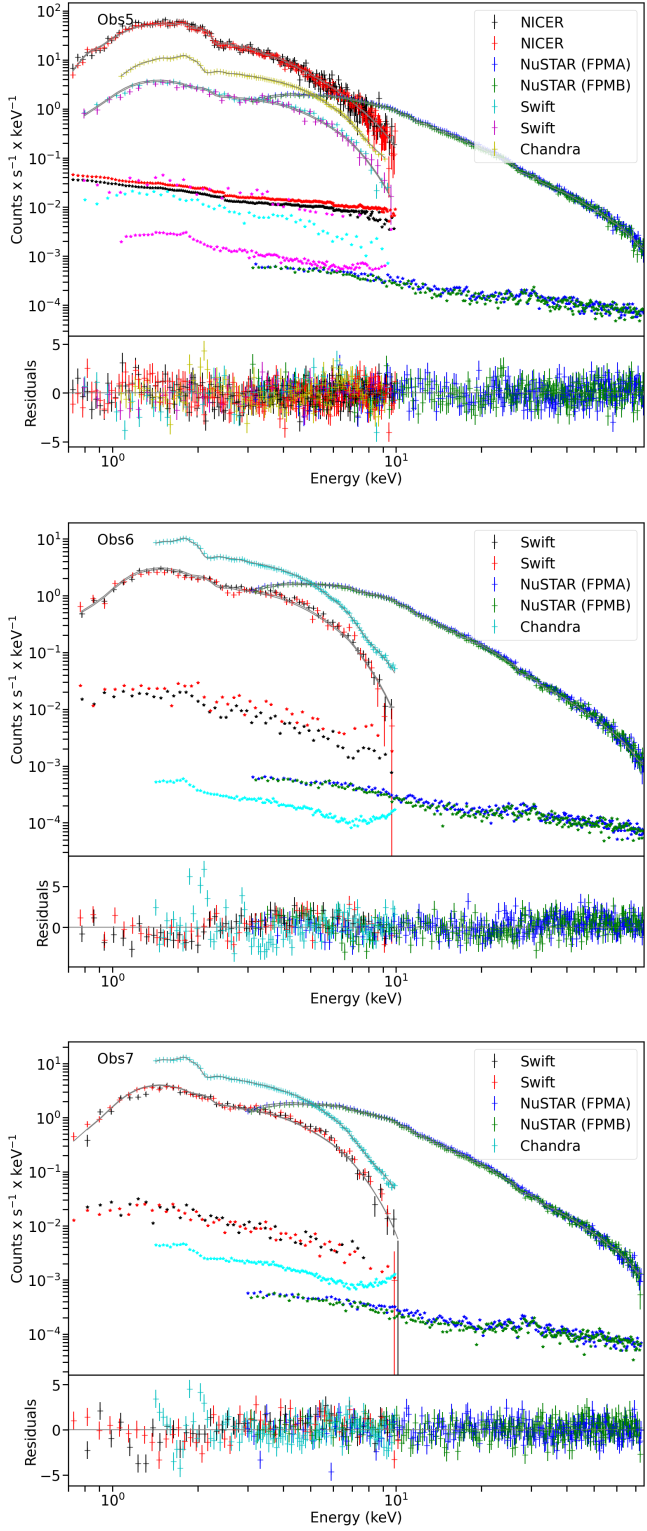

**Figure 8.** *Top*: *Chandra* in yellow, jointly with *NICER*, *Swift* and *NuSTAR*, corresponding to observation 5. *Middle*: *Chandra* data is represented by the color cyan jointly with *Swift* and *NuSTAR*. *Bottom*: *Chandra* data is represented by the color cyan jointly with *Swift* and *NuSTAR*. In all cases, *Chandra* is consistent with the other instruments.